\documentclass[a4paper,11pt]{article}
\usepackage[T1]{fontenc} \usepackage{lmodern}  
\usepackage{a4wide}
\usepackage{amssymb,amsmath}
\usepackage{mathtools}  
\usepackage{color}
\usepackage{graphicx}  
\usepackage{footmisc}  

\usepackage{url}   
\usepackage{fancyhdr}  

\usepackage{hyperref}   
\hypersetup{hidelinks, colorlinks=true, linkcolor=blue, citecolor=blue}

\DeclareMathOperator*{\argmax}{arg\,max}

\begin{document}

\title{Assessment of COVID-19 hospitalization forecasts from a simplified SIR model\footnotemark[1]}

\author{P.-A. Absil\footnotemark[4] \footnotemark[3] \and Ousmane Diao\footnotemark[4] \and Mouhamadou Diallo\footnotemark[5]}

\maketitle

\renewcommand{\thefootnote}{\fnsymbol{footnote}} 
\footnotetext[1]{The second author is supported by a fellowship awarded by UCLouvain's Conseil de l'action internationale.
} 
\footnotetext[4]{ICTEAM Institute, UCLouvain, B-1348 Louvain-la-Neuve, Belgium}
\footnotetext[5]{Molecular Biology Unit/Bacteriology-Virology Lab, CNHU A. Le Dantec / Universit\'e Cheikh Anta Diop, Dakar, S\'en\'egal}
\footnotetext[3]{Corresponding author - email: absil@inma.ucl.ac.be - url: \url{http://sites.uclouvain.be/absil/}}
\renewcommand{\thefootnote}{\arabic{footnote}}


\thispagestyle{fancy}
\lhead{Tech.\ report UCL-INMA-2020.05-v2}
\rhead{\url{http://sites.uclouvain.be/absil/2020.05}}

\bigskip

\begin{abstract}
We propose the SH model, a simplified version of the well-known SIR compartmental model of infectious diseases. With optimized parameters and initial conditions, this time-invariant two-parameter two-dimensional model is able to fit COVID-19 hospitalization data over several months with high accuracy (e.g., the root relative squared error is below 10\% for Belgium over the period from 2020-03-15 to 2020-07-15).
Moreover, we observed that, when the model is trained on a suitable three-week period around the first hospitalization peak for Belgium, it forecasts the subsequent two months with mean absolute percentage error (MAPE) under 4\%.
We repeated the experiment for each French department and found 14 of them where the MAPE was below 20\%.
However, when the model is trained in the increase phase, it is less successful at forecasting the subsequent evolution.
\end{abstract}


\section{Introduction}

Compartmental models, and in particular the well-known SIR model~\cite{KermackMcK1927}, have been widely used to model infectious diseases since the early 20th century; see~\cite{Hethcote2000} and references therein.
Letting $S(t)$, $I(t)$, and $R(t)$ denote the number of susceptible, infectious and removed (or recovered) individuals at time $t$, and letting $\dot S(t)$, $\dot I(t)$, and $\dot R(t)$ denote their time derivatives, the SIR model consists in the following three-dimensional continuous-time autonomous dynamical system
\begin{subequations}  \label{eq:SIR}
\begin{align}
  \dot S(t) &= - \frac{\beta}N S(t) I(t)  \label{eq:S}
  \\ \dot I(t) &= \frac{\beta}N S(t) I(t) - \gamma I(t)  \label{eq:I}
  \\ \dot R(t) &= \gamma I(t)  \label{eq:R},
\end{align}
\end{subequations}
where $N = S(t) + I(t) + R(t)$ is the constant total population and $\beta$ and $\gamma$ are parameters.
The SIR model, and several (sometimes deep) variations thereof, have been applied in several works to model the COVID-19 dynamics (see, e.g.,~\cite{LiuGayWilRoc2020,Atkeson2020,2003.14391,FanelliPia2020,CarlettiFanPia2020,KOZYREFF2021398,Nesterov2020,Nesterov2007.11429,Djidjou-Demasse2020.04.02.20049189,Abrams2021}) with known limitations~\cite{RodaVarHanLi2020,2006.06373,WangFle2020,ComunianGabGiu2020}. 
In~\cite{BhanotDeL2020}, an SIR-like model is used to make long-term forecasts. 
However, at the time of writing this paper, it appears that studies are still rare (see~\cite{Singh2020.07.08.20148619,GhoshCha2020,GhanamBooAbd2020}) where the SIR-like model parameters and initial conditions are learned on a ``train'' part of the available data in order to predict a subsequent ``test'' part, making it possible to assess the forecast accuracy of the model.

In this paper, we adapt the SIR model to the situation where (i) $S$, $I$ and $R$ are hidden variables but $I(t)$ is observed through a ``proxy'' $H(t) = \alpha I(t)$, where $\alpha$ is unknown but constant, and (ii) not only $\beta$ and $\gamma$ but also the total concerned population $N$ are unknown.
In the context of the COVID-19 application, the proxy $H$ will be the total number of lab-confirmed hospitalized patients; the equation $H(t) = \alpha I(t)$ thus posits that a constant fraction of the infected people is hospitalized. 
The proposed adapted SIR model, which we term \emph{SH model}, is given in~\eqref{eq:SH}. It has two state variables ($\bar{S}$---a scaled version $S$---and $H$) and two parameters ($\bar\beta$---which lumps together the parameters $\beta$, $N$, and $\alpha$---and $\gamma$). 

We leverage the proposed SH model as follows in order to make hospitalization forecasts. Given observed values $(H_o(t))_{t=t_i,\dots,t_c}$, we estimate the parameters $\bar\beta$, $\gamma$, and the initial conditions $\bar{S}(t_i)$ and $H(t_i)$ of the SH model. Then we simulate the SH model in order to predict $(H(t))_{t=t_c+1,\dots,t_f}$ for a specified final prediction time $t_f$. This approach thus 
relates to
the areas of parameter estimation (for obvious reasons), data assimilation (for the generation of the initial conditions) and machine learning (for the train-test approach).

The paper is organized as follows. The available data is reviewed in Section~\ref{sec:data}. The SH model is derived in Section~\ref{sec:models}. Several methods to estimate the parameters and initial conditions are presented in Sections~\ref{sec:estimation} and~\ref{sec:alternatives}. Experimental results are reported in Section~\ref{sec:results} and conclusions are drawn in Section~\ref{sec:conclusion}.

\section{Data and notation}
\label{sec:data}

In the experiments (Section~\ref{sec:results}), we will use a COVID-19 dataset for Belgium\footnote{\url{https://epistat.sciensano.be/Data/COVID19BE_HOSP.csv} obtained from \url{https://epistat.wiv-isp.be/covid/}} that provides us with the following data for $t=t_s,\dots,t_e$, where $t_s$ is 2020-03-15 and $t_e$ is 2021-05-15:
\begin{itemize}
\item $H_o(t)$: total number of lab-confirmed hospitalized (i.e., at the hospital) COVID-19 patients on day $t$, including intensive care unit
(TOTAL\_IN column); 
\item $E_o(t)$: number of COVID-19 patients entering the hospital (number of lab-confirmed hospital intakes) on day $t$ (NEW\_IN column); 
\item $L_o(t)$: number of COVID-19 patients discharged from the hospital on day $t$ (NEW\_OUT column).
\end{itemize}
The subscript $_o$ stands for ``observed''.

We will also mention results obtained with a dataset for France\footnote{donnees-hospitalieres-covid19-2021-05-16-19h08.csv obtained from \url{https://www.data.gouv.fr/en/datasets/donnees-hospitalieres-relatives-a-lepidemie-de-covid-19/}. Department 46 was removed from the data due to abnormal values in the ``rad'' column.} where $t_s$ is 2020-03-18 and $t_e$ is 2021-05-16.

\subsection{Discussion}
In the data for Belgium, there is a mismatch between $H_o(t)$ and $H_o(t-1) + E_o(t) - L_o(t)$ for most $t$, 
and $H_o(t_s) + \sum_{t=t_s+1}^{t_e} E_o(t)-L_o(t)$ is significantly larger than $H_o(t_e)$. This can be due to the patients who get infected at the hospital (they would be counted in $H_o$ without appearing in $E_o$) and to the patients who die at the hospital (they would be removed from $H_o$ without appearing in $L_o$). 
In order to remedy this mismatch, we redefine $L_o(t)$ by $L_o(t) := -H_o(t) + H_o(t-1) + E_o(t)$. 

For the French data, we sum the ``rad'' (daily number of new home returns) and ``dc'' (daily number of newly deceased persons) columns to get $L_o(t)$. Since there is no column for $E_o$, we define $E_o(t) = H_o(t)-H_o(t-1) + L_o(t)$.

Several other COVID-19-related data are available for Belgium and France. In particular, the daily number of infected individuals, $I_o(t)$, is also reported by health authorities. However, the decision process to admit individuals at the hospital is believed to be more stable over time than the decision process to test individuals, hence the hospitalization data is a priori more prone to be modeled by a dynamical system with time-invariant parameters.
Moreover, for the authorities, predicting $H$ is more crucial than predicting $I$.
Therefore, as in~\cite{KOZYREFF2021398}, we focus on $H$.

\section{Models}
\label{sec:models}

\subsection{Case hospitalization ratio}
\label{sec:case-hosp-ratio}

We assume that, for all $t$,
\begin{equation}  \label{eq:alpha}
  H(t) = \alpha I(t)
\end{equation}
where $\alpha$ is unknown but constant over time. As already mentioned,~\eqref{eq:alpha} posits that a constant fraction of the infected people is hospitalized. 

Equation~\eqref{eq:alpha} is reminiscent of~\cite[(3)]{CarlettiFanPia2020}, where the number of dead individuals plays the role of $H$ and $\alpha$ is time dependent.

\subsection{Observation models}

We assume the following observation models with additive noise:
\begin{subequations}  \label{eq:observed}
\begin{align}
  H_o(t) &= H(t) + \epsilon_H(t)  \label{eq:Ho}
  \\ E_o(t) &= E(t) + \epsilon_E(t)    
  \\ L_o(t) &= L(t) + \epsilon_L(t).  
\end{align}
\end{subequations}
The variables with the subscript $_o$ stand for the values provided in the datasets; see Section~\ref{sec:data}. The variables without the subscript stand for the values given by the forthcoming model. The $\epsilon$ variables account for the discrepancies between the former and the latter. 

We do not make any explicit statistical assumption on the $\epsilon$ variables in this work. However, we point out that if we (questionably) assume them to be independent Gaussian centered random variables, then some subsequent estimators admit a maximum likelihood interpretation.

\subsection{Proposed SH model}

Multiplying~\eqref{eq:S} and~\eqref{eq:I} by $\alpha$, and further multiplying the numerators and denominators by $\alpha$, we obtain the equivalent system
\begin{align}
  \alpha \dot{S}(t) &= -\frac{\beta}{N\alpha} \, \alpha S(t) \, \alpha I(t)
  \\ \alpha \dot{I}(t) &= \frac{\beta}{N\alpha} \, \alpha S(t) \, \alpha I(t) - \gamma \alpha I(t).
\end{align}
Letting
\begin{align}
  \bar{S} &:= \alpha S  \label{eq:barS}
  \\ \bar\beta &:= \frac{\beta}{N\alpha}  \label{eq:barbeta}
\end{align}
and using~\eqref{eq:alpha}, we obtain the simplified SIR model
\begin{subequations}  \label{eq:SH}
\begin{align}
  \dot{\bar{S}}(t) &= -\bar\beta \bar{S}(t) H(t)  \label{eq:SH-S}
  \\ \dot{H}(t) &=  \bar\beta \bar{S}(t) H(t) - \gamma H(t)  \label{eq:SH-H}
\end{align}
\end{subequations}
which we term the \emph{SH model}. (The ``S'' in this SH model can be interpreted as the number of individuals susceptible of being hospitalized.) The SH model has only two parameters ($\bar\beta$ and $\gamma$), one hidden state variable ($\bar{S}$) and one observed state variable ($H$) with observation model~\eqref{eq:Ho}.

Note that, in the SH model~\eqref{eq:SH}, the number of patients entering the hospital by unit of time is
\begin{equation}  \label{eq:E}
E(t) := \bar\beta \bar{S}(t) H(t)
\end{equation}
and the number of patients leaving the hospital by unit of time is
\begin{equation}  \label{eq:L}
L(t) := \gamma H(t).
\end{equation}

\section{Estimation and prediction method}
\label{sec:estimation}

The goal is now to leverage the SH model~\eqref{eq:SH} in order to predict future values of $H$ based on its past and current observations $(H_o(t))_{t=t_s,\dots,t_c}$. To this end, we have to estimate (or ``learn'') four variables, which we term \emph{estimands}: the two parameters $\bar\beta$ and $\gamma$ and the two initial values $\bar{S}(t_i)$ and $H(t_i)$, where $t_i$ is the chosen initial time for the SH model~\eqref{eq:SH}. One possible approach is to minimize some error measure between the simulated values $(H(t))_{t=t_i,\dots,t_c}$ and the observed values $(H_o(t))_{t=t_i,\dots,t_c}$ as a function of the four estimands. However, the error measure is not available as a closed-form expression of the four estimands, and this makes this four-variable optimization problem challenging. We show in this section that it is possible to estimate $H(t_i)$ and $\gamma$ separately. This leaves us with an optimization problem in the two remaining estimands $\bar\beta$ and $\bar{S}(t_i)$, making it possible to visualize the objective function by means of a contour plot. 

\subsection{Train and test sets}

To recap, we have $t_s \leq t_i < t_c < t_e$. The provided dataset goes from $t_s$ to $t_e$. The \emph{test set}, $(H_o(t),E_o(t),L_o(t))_{t\in[t_c+1,t_e]}$, is not available to the forecasting algorithm; it is revealed only to compute the forecasting error (see Section~\ref{sec:accuracy}). The SH model is initialized at $t_i$, and we refer to the data $(H_o(t),E_o(t),L_o(t))_{t\in[t_i,t_c]}$ as the \emph{train set}, though it is legitimate to widen it to $t\in[t_s,t_c]$.

\subsection{Estimation of $H(t_i)$}
\label{sec:H0}

We simply take
\[
H(t_i) := H_o(t_i).
\]
In view of~\eqref{eq:Ho}, this amounts to $\epsilon_H(t_i)=0$.

We could have chosen instead $H(t_i)$ as a filtered version of $H_o(t_i)$ in order to reduce the impact of the weekly variations in $H_o(t)$. Those weekly variations, visible in Figure~\ref{fig:SHR_12PA_BEL_traintstart1_traintstop117_c100_4Dopt} in the form of equally spaced ripples, are due to the fact that fewer patients are discharged during the weekend. However, such a filter did not yield better forecasts in our experiments, and we thus opted for the above-mentioned $H(t_i) := H_o(t_i)$.

\subsection{Estimation of $\gamma$}
\label{sec:gamma}

We have $L(t) = \gamma H(t)$, see~\eqref{eq:L}.
In view of the observation model~\eqref{eq:observed}, we can estimate $\gamma$ by a ratio of means:
\begin{equation}  \label{eq:gamma}
\hat\gamma^{\text{RM}} = \frac{\sum_{t=t_i}^{t_c} L_o(t)}{\sum_{t=t_i}^{t_c} H_o(t)}.
\end{equation}
A theoretical justification of the choice of the ratio-of-means estimator can be found in~\cite{SchoukensPintelon1991}; or see~\cite[\S 1.2]{Schoukens2018}.

Note that $t_i$ in the expression of $\hat\gamma$ can legitimately be replaced by any time between $t_s$ and $t_c$. Only data in the test set, i.e., occurring after $t_c$, are unavailable in the variable estimation phase.

\subsection{Joint estimation of $\bar\beta$ and $\bar{S}(t_i)$}
\label{sec:b-S0}

Now we have to estimate the two remaining estimands, namely $\bar\beta$ and $\bar{S}(t_i)$. We choose the following sum-of-squared-errors objective function
\begin{equation}  \label{eq:phi}
\phi(\bar\beta,\bar{S}(t_i)) = c_H \sum_{t=t_i}^{t_c} (H(t) - H_o(t))^2 + c_E \sum_{t=t_i}^{t_c} (E(t) - E_o(t))^2 + c_L \sum_{t=t_i}^{t_c} (L(t) - L_o(t))^2,
\end{equation}
where the $c$ coefficients are parameters, all set to $1$ in our experiments unless otherwise stated.
In~\eqref{eq:phi}, $H(t)$, $E(t)$ (see~\eqref{eq:E}), and $L(t)$ (see~\eqref{eq:L}) are given by the (approximate) solution of the SH model~\eqref{eq:SH} in which (i) $H(t_i)$ and $\gamma$ take the values estimated as above, and (ii) $\bar\beta$ and $\bar{S}(t_i)$ take the values specified in the argument of $\phi$. 

Several methods exist to compute the required (approximate) solution of the SH model~\eqref{eq:SH}; see~\cite{BarlowWei2020}. Since the method will have to be repeatedly called by the optimization solver, we opt for the simplicity of the explicit Euler method with a time step of one day (as in, e.g.,~\cite{Vinitsky2021}), yielding, for $t=t_i,\dots,t_c-1$,
\begin{subequations}  \label{eq:SH-DT}
\begin{align}
  \bar{S}(t+1) &= \bar{S}(t) - \bar\beta \bar{S}(t) H(t)  \label{eq:SH-DT-S}
  \\ H(t+1) &= H(t) + \bar\beta \bar{S}(t) H(t) - \gamma H(t).  \label{eq:SH-DT-H}
\end{align}
\end{subequations}

As an alternative numerical integration method for~\eqref{eq:SH}, we tested SciPy's explicit Runge--Kutta method of order 5(4) with default parameters. In comparison with Euler~\eqref{eq:SH-DT}, the forecast error was most often in favor of the Runge--Kutta method, but always by a narrow margin, and the execution time with Runge-Kutta was more than four times the one with Euler. For these reasons, we decided to retain~\eqref{eq:SH-DT}. This choice also leads to simpler developments in Section~\ref{sec:b-then-S0}.

Now that the objective function $\phi$ (also termed ``cost function'' or ``loss function'') is defined, we let the estimated $(\bar\beta,\bar{S}(t_i))$ be the (approximate) minimizer of $\phi$ returned by some optimization solver. 
The initial guess that we give to the solver is the outcome of the successive estimation method that we will present in Section~\ref{sec:b-then-S0}. (If this proposed initial guess is negative, a situation that we observed on rare occasions in the late part of the decrease phase, then we take its opposite.)

\subsection{Prediction of $H$}

Recall that the time range between $t_i$ and $t_c$ is the train period and the time range between $t_c+1$ and $t_e$ is the test period.
In order to predict (i.e., forecast) the values of $H$ over the test period, we apply the above procedure to estimate the four estimand variables $\bar\beta$, $\gamma$, $\bar{S}(t_i)$, and $H(t_i)$, and we compute the solution $H(t)$ of~\eqref{eq:SH-DT} for $t$ from $t_i$ to $t_e$. The prediction is then $(H(t))_{t=t_c+1,\dots,t_e}$.
The discrepancy between $(H(t))_{t=t_c+1,\dots,t_e}$ and $(H_o(t))_{t=t_c+1,\dots,t_e}$ reveals the accuracy of the prediction.

\section{Alternative estimation and prediction methods}
\label{sec:alternatives}

\subsection{Successive estimation of $\bar\beta$ and $\bar{S}(t_i)$}
\label{sec:b-then-S0}

As an alternative to Section~\ref{sec:b-S0}, we now present a method to estimate $\bar\beta$ independently. We do not recommend this alternative, as we have observed that it usually yields less accurate forecasts. However, it provides a convenient initial guess for Section~\ref{sec:b-S0}, and it also sheds light on the various forecast accuracies observed in Section~\ref{sec:results}.  

From~\eqref{eq:SH-DT-S} and~\eqref{eq:E}, we obtain
\[
\frac{E(t+1)}{H(t+1)} - \frac{E(t)}{H(t)} = - \bar\beta E(t).
\]
Summing both sides yields the following estimator of $\bar\beta$:
\begin{equation}  \label{eq:hat-bar-beta-easy-RM}
\widehat{\bar\beta} = - \frac{ \frac{E_o(t_c)}{H_o(t_c)}-\frac{E_o(t_i)}{H_o(t_i)} }{\sum_{t=t_i}^{t_c-1} E_o(t)}.
\end{equation}
Finally, in view of~\eqref{eq:E}, we choose the following estimator for $\bar{S}(t_i)$:
\[
\widehat{\bar{S}}(t_i) = \frac{E_o(t_i)}{\widehat{\bar\beta} H_o(t_i)}.
\]

For the observation model~\eqref{eq:observed}, under the assumption that $|\epsilon_H(t)|\leq \epsilon$ and $|\epsilon_E(t)|\leq \epsilon$ for all $t$, we have the following error bound, where higher order terms in $\epsilon$ are neglected:
\[
\left| \frac{ \frac{E_o(t_c)}{H_o(t_c)} }{\sum_{t=t_i}^{t_c-1} E_o(t)} - \frac{ \frac{E(t_c)}{H(t_c)} }{\sum_{t=t_i}^{t_c-1} E(t)} \right| \lesssim \frac{\epsilon}{H(t_c) \sum_{t=t_i}^{t_c-1} E(t)} \left( 1+\frac{E(t_c)}{H(t_c)} + \frac{E(t_c) (t_c-t_i)}{\sum_{t=t_i}^{t_c-1} E(t)}  \right),
\] 
and likewise for the second term in~\eqref{eq:hat-bar-beta-easy-RM}.

Consequently, the estimation of $\bar\beta$ should be the most accurate when $E_o(t)H_o(t)$ is the largest. This occurs slightly before the peak of $H_o(t)$. This means that the estimation of $\bar\beta$ should be the most accurate for a train period slightly before the peak. However, this does not mean that this position of the train period gives the most accurate forecasts, as we will see below.

Let us consider the situation where the train period is located \emph{before} the peak. Then the estimation of $\bar\beta$ is less accurate, and this impacts $\widehat{\bar{S}}(t_i)$. At the initial time $t_i$, this does not impact the right-hand term of~\eqref{eq:SH-H} in view of the definition of $\widehat{\bar{S}}(t_i)$. However, an overestimation of $\bar\beta$ will induce an underestimation of $\bar{S}(t_i)$ and, in view of~\eqref{eq:SH-S}, a subsequent even stronger underestimation of $\bar{S}(t)$. Hence the first term of~\eqref{eq:SH-H} will be underestimated. As a consequence, the peak in $H$ will appear sooner and lower. The case of an underestimation of $\bar\beta$ leads to the opposite conclusion, namely a peak in $H$ that appears later and higher. In summary, the further before the peak the train period is located, the more inaccurate the position and height of the peak is expected to be.

Finally, let us consider the situation where the train period is located \emph{after} the peak. Then we can make the same observations as in the previous paragraph, except that predicting the peak is now irrelevant. Moreover, we are in the decrease phase, where the first term of~\eqref{eq:SH-H} (which involves $\bar\beta$ and $\bar{S}(t)$) is smaller than the second term (which does not involve these quantities). Consequently, the possibly large estimation errors on $\bar\beta$ and $\bar{S}(t)$ will only slightly affect the forecast of $H(t)$.

\subsection{Joint estimation of the four estimands}
\label{sec:all-estimands}

An alternative to Sections~\ref{sec:H0}--\ref{sec:b-S0} is to reconsider~\eqref{eq:phi} as a function of all four estimands:
\begin{equation}  \label{eq:phitilde}
\tilde\phi(\bar\beta,\bar{S}(t_i),\gamma,H(t_i)) = c_H \sum_{t=t_i}^{t_c} (H(t) - H_o(t))^2 + c_E \sum_{t=t_i}^{t_c} (E(t) - E_o(t))^2 + c_L \sum_{t=t_i}^{t_c} (L(t) - L_o(t))^2.
\end{equation}
In~\eqref{eq:phitilde}, $H(t)$, $E(t)$ (see~\eqref{eq:E}), and $L(t)$ (see~\eqref{eq:L}) are given by the solution of the discrete-time SH model~\eqref{eq:SH-DT} where the parameters $\bar\beta$ and $\gamma$ and the initial conditions $\bar{S}(t_i)$ and $H(t_i)$ take the values specified in the argument of $\tilde\phi$. Minimizing $\tilde\phi$ is a more challenging problem than minimizing $\phi$~\eqref{eq:phi} in view of the larger number of optimization variables. It may be essential to give a good initial guess to the optimization solver, and a natural candidate for this is the values obtained by the procedure described in Sections~\ref{sec:H0}--\ref{sec:b-S0}.

In our 
experiments, we have found that this alternative does not present a clear advantage in terms of the prediction mean absolute percentage error (MAPE). The results reported in Section~\ref{sec:results} are obtained with the prediction approach of Section~\ref{sec:estimation}, unless otherwise specified.

\section{Results}  
\label{sec:results}

We now apply the method of Section~\ref{sec:estimation} (by default) or Section~\ref{sec:all-estimands} (when specified) to the data of Section~\ref{sec:data}.

The methods are implemented in Python 3 and run with Anaconda 2019.10. 
The code to reproduce the results is available from \url{https://sites.uclouvain.be/absil/2020.05}.

\subsection{Accuracy measures}
\label{sec:accuracy}

We will conduct experiments for more than 100 districts with various train and test periods. Since plotting all the results is out of the question, we have to return measures that give an account of the accuracy of the predictions. Among the many possibilities, we will favor the measures defined as follows for an arbitrary period from $t_a$ to $t_b$:
\begin{itemize}
\item Root mean square error (RMSE):
\begin{equation}  \label{eq:RMSE}
\textrm{RMSE} = \sqrt{ \frac{\sum_{t=t_a}^{t_b}(H_o(t) - H(t))^2}{t_b-t_a+1} }.
\end{equation}
The numerator corresponds to the first term of the objective function~\eqref{eq:phi}. From an optimization viewpoint, it has the advantage of being a smooth function.

\item Root relative squared error (RRSE):
\begin{equation}
\textrm{RRSE} = \sqrt{ \frac{\sum_{t=t_a}^{t_b}(H_o(t) - H(t))^2}{\sum_{t=t_a}^{t_b}(H_o(t) - \textrm{mean}(H_o))^2} }
\end{equation}
where $\textrm{mean}(H_o) = \frac{1}{t_b-t_a+1} \sum_{t=t_a}^{t_b} H_o(t)$. This measure can be interpreted as the ratio between the RMSE of $H$ and the RMSE of $\textrm{mean}(H_o)$. 

\item Mean absolute error (MAE):
\begin{equation}
\textrm{MAE} = \frac1{t_b-t_a+1} \sum_{t=t_a}^{t_b}\left|H_o(t)-H(t)\right|.
\end{equation}
The MAE is considered to have a better interpretability than the RMSE. However, for comparisons between datasets, scale-invariant versions, given next, should be favored.

\item Mean absolute scaled error (MASE):
\begin{equation}  \label{eq:MASE}
\textrm{MASE} = \frac{\textrm{MAE}}{\frac1{t_b-t_a} \sum_{t=t_a}^{t_b-1}|H_o(t+1)-H_o(t)|}.
\end{equation}
This measure is inspired from~\cite{HyndmanKoe2006}. It consists of the ratio between the MAE and the mean daily variation of the observed values. An MASE value around 1 or below indicates an excellent accuracy.

\item Mean absolute percentage error (MAPE):
\begin{equation}
\textrm{MAPE} = \frac1{t_b-t_a+1} \sum_{t=t_a}^{t_b} \frac{|H_o(t)-H(t)|}{|H_o(t)|}.
\end{equation}
While easy to interpret, this measure may be considered to give an undue importance to errors on small values. Moreover, in the case, encountered in practice, where $H_o(t)=0$ for some $t$, it is undefined. 

\item Symmetric mean absolute percentage error (sMAPE):
\begin{equation}
\textrm{sMAPE} = \frac1{t_b-t_a+1} \sum_{t=t_a}^{t_b} \frac{|H_o(t)-H(t)|}{(|H_o(t)|+|H(t)|)/2}.
\end{equation}
This measure overcomes the undefinedness issue of the MAPE since $H(t)$ is always (strictly) positive in the SH model.
\end{itemize}

\subsection{Fitting experiment}

\begin{figure}
\centerline{
\includegraphics[width=.5\textwidth]{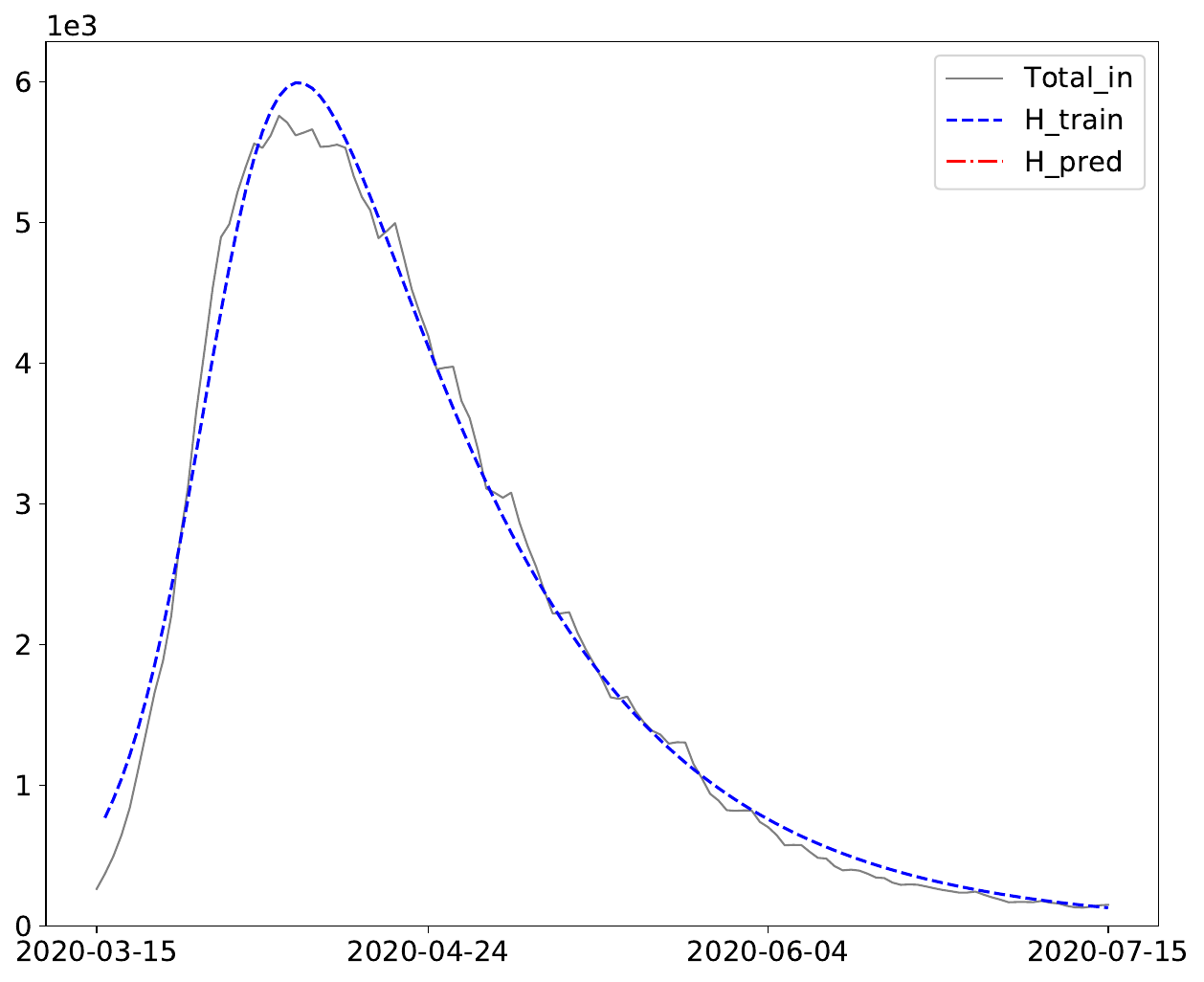}
\includegraphics[width=.5\textwidth]{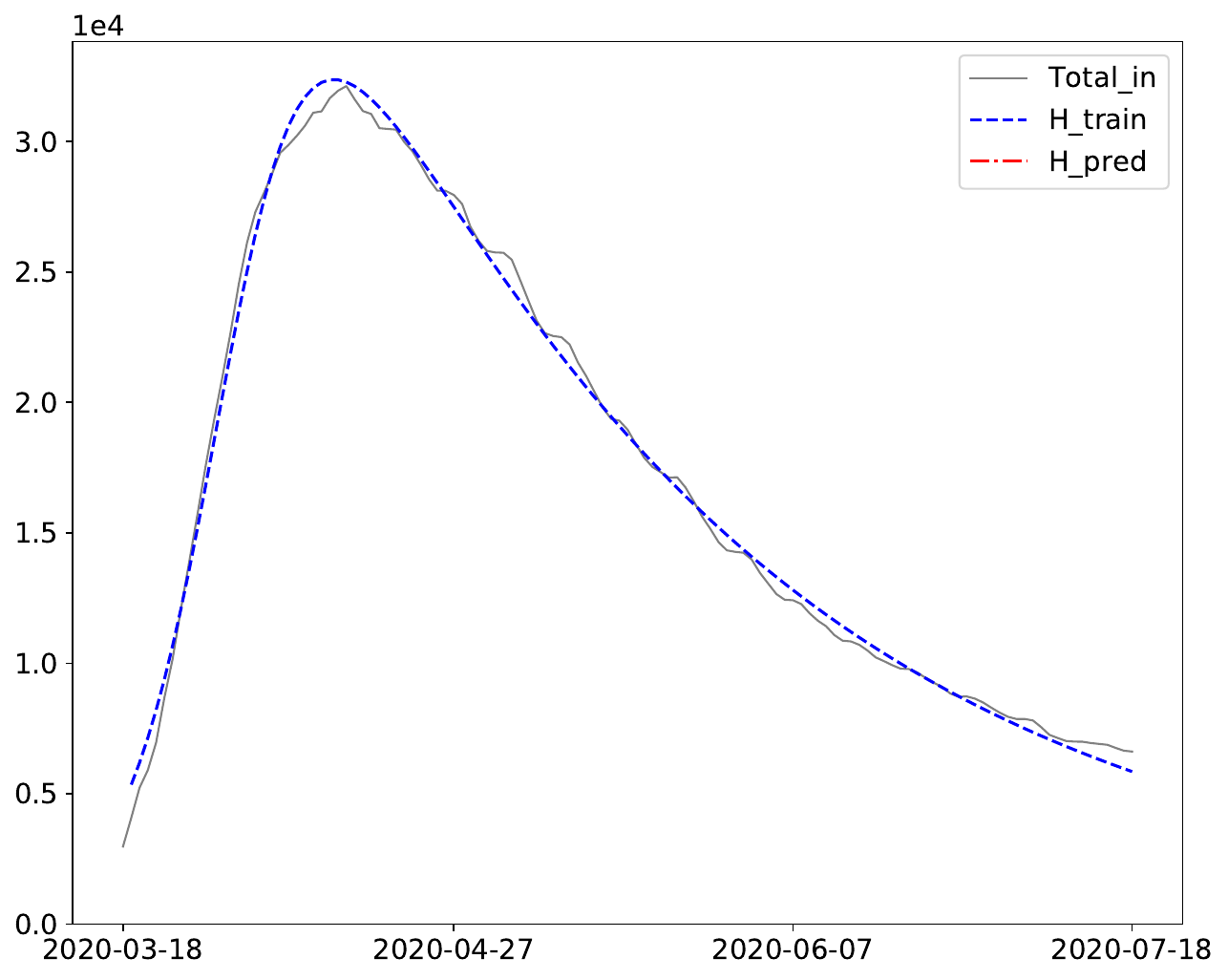}
}
\caption{Fitting the SH model to the $H_o$ (total hospitalized) curve. In this experiment, the train set is the whole dataset, hence there is no test (i.e., forecast) curve. Left: Belgium. Right: France.}
\label{fig:SHR_12PA_BEL_traintstart1_traintstop117_c100_4Dopt}
\end{figure}

We first check how well the SH model~\eqref{eq:SH-DT} can fit the available data. For this experiment, we use the method of Section~\ref{sec:all-estimands} with $c_E=c_L=0$ in order to get the best possible fit (in the least squares sense) to the $H_o$ time series. The results are shown in Figure~\ref{fig:SHR_12PA_BEL_traintstart1_traintstop117_c100_4Dopt} for Belgium\footnote{Reproduce with SHR\_22PA\_py\_BELsum\_1sttraintstart1\_1sttraintend123\_1sttesttend123\_c111.zip} and France.\footnote{Reproduce with SHR\_22PA\_py\_FRAsum\_1sttraintstart1\_1sttraintend123\_1sttesttend123\_c111.zip} 
In this figure, the data is restricted to the first four months, where only one wave is present. Note that a tight fit over a period containg more than one wave is unachievable since the solutions of model~\eqref{eq:SH-DT} have only one peak.

For Belgium, the RRSE is 9\% and the MAPE 14\%. For France, the RRSE is 6\% and the MAPE 3\%. The comparatively large value of the MAPE for Belgium illustrates the comment made about MAPE in Section~\ref{sec:accuracy}.

The fitting error is remarkably small in view of the fact that there are some 120 data points for only 4 estimands. Recall indeed that the parameters of the SH model are constant with respect to time in our experiments. This contrasts with~\cite{KOZYREFF2021398} where there are two phases, and with~\cite{Nesterov2020} where the infection rate is piecewise constant with several pieces.

We stress that Figure~\ref{fig:SHR_12PA_BEL_traintstart1_traintstop117_c100_4Dopt} in itself does not imply that the model leads to accurate \emph{forecasts}. If the optimal fit over some period is bad, then forecasts over that period can only be bad. But if the fit is good (as it is the case here), forecasts can still be bad due to their sensitivity with respect to the data preceding the to-be-forecasted period. For example, a better fit (in the RMSE sense) than in Figure~\ref{fig:SHR_12PA_BEL_traintstart1_traintstop117_c100_4Dopt} (left) can be obtained with a polynomial of degree 8; however, its forecast accuracy is abysmal.

In order to assess the forecast accuracy of the model, we have to learn the estimand variables over a \emph{train period} that we make available to the algorithm, then use the learned estimands in order to predict $H$ over a subsequent \emph{test period} whose data is not available to the algorithm, and finally compare the prediction with the data on the test period. This is what we proceed to do in the rest of this Section~\ref{sec:results}.

\subsection{Forecasts from various train periods}
\label{sec:all-train-periods}

Since the solutions of the SH model~\eqref{eq:SH-DT} have only one peak, it is already known that its forecast accuracy will be poor if the train and test periods overlap with two successive waves. In this section, we investigate if there exist train periods such that the subsequent few weeks are forecasted with high accuracy by the SH model.

\begin{figure}[t]
\centerline{
\includegraphics[width=.9\textwidth]{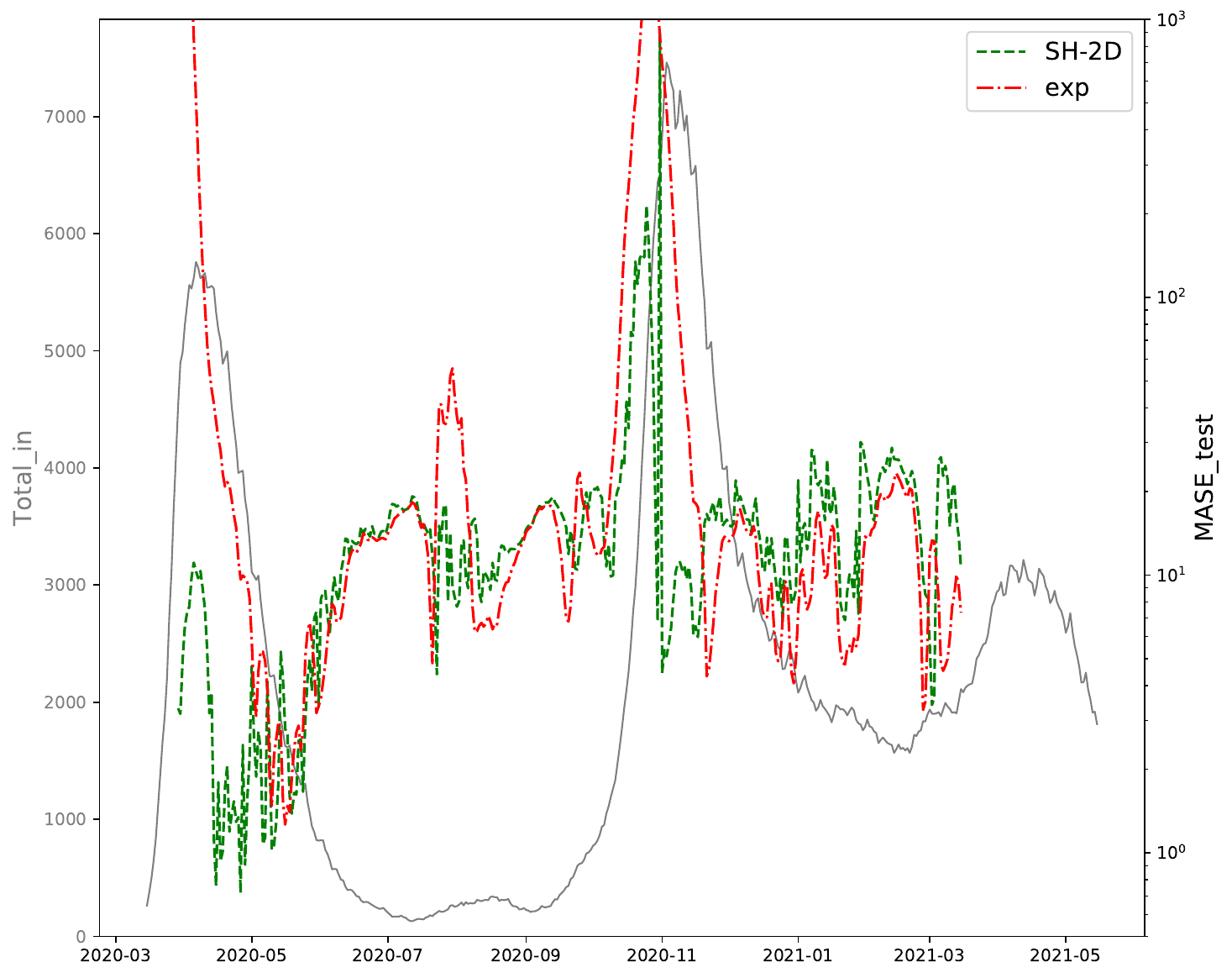}
}
\caption{Forecast accuracy assessment for Belgian data. As a function of $t$ (horizontal axis), the figure shows $(H_o(t))_{t=t_s,\dots,t_e}$ (observed hospitalizations, gray solid line, left-hand vertical axis), and the MASE over the test period $[t+1,t+60]$ for the train period $[t-13,t]$ obtained respectively with the SH model~\eqref{eq:SH-DT} using the estimation method of Section~\ref{sec:estimation} (green dashed line, right-hand vertical axis) and with the exponential model presented at the end of Section~\ref{sec:all-train-periods} (red dash-dotted line, right-hand vertical axis).}
\label{fig:SHR_22PA_py_BELsum_1sttraintstart1_1sttraintend15_1sttesttend75_c111_MASE}
\end{figure}

For the Belgian data, Figure~\ref{fig:SHR_22PA_py_BELsum_1sttraintstart1_1sttraintend15_1sttesttend75_c111_MASE}\footnote{Reproduce with SHR\_22PA\_py\_BELsum\_1sttraintstart1\_1sttraintend15\_1sttesttend75\_c111.zip} shows, as a function of $t_c$, the MASE over the 60-day period $[t_c+1,t_c+60]$ obtained with the two-week train period $[t_c-13,t_c]$. As anticipated, for some train periods, the forecast error is high, with an MASE around 100 or above. For most train periods, the MASE is around 10. However, when the train period is located around the first peak, the MASE goes down to around 1. This means that the forecast error over the 60 days is so low that it is comparable to the daily variation of the data. 

\begin{figure}
\centerline{
\includegraphics[width=.9\textwidth]{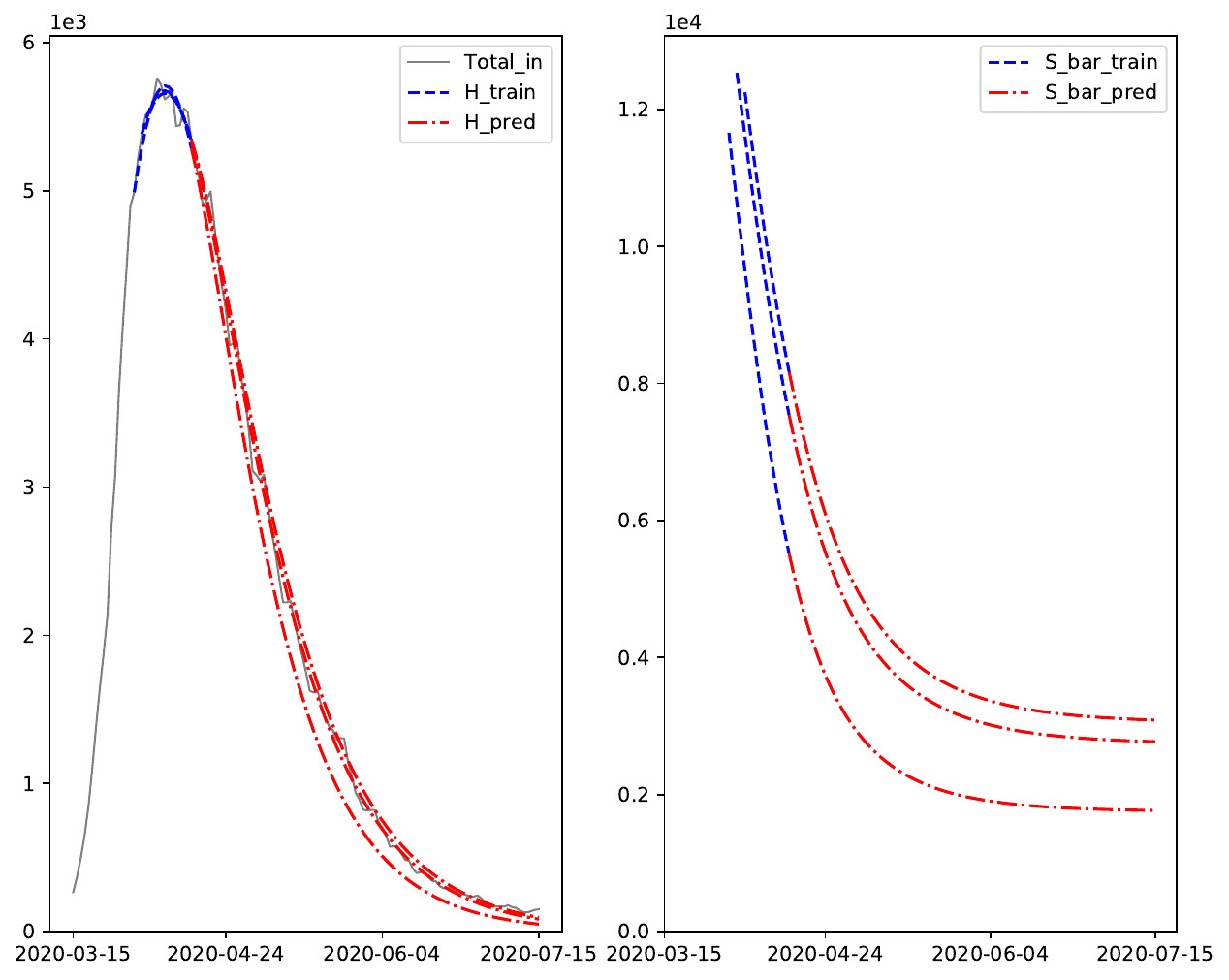}}
\caption{Belgium, train periods around the first peak. 
The left-hand plot shows $(H_o(t))_{t=t_s,\dots,t_e}$ (observed hospitalizations, gray solid line), $(H(t))_{t=t_i,\dots,t_c}$ (hospitalizations given by the model over the train period, blue dashed line), and $(H(t))_{t=t_c+1,\dots,t_e}$ (hospitalizations forecasted over the test period, in red dash-dot line). In order to give a sense of the sensitivity of the results, we superpose the curves obtained for three slightly different train periods. The test MASE values for the three curves are 3.16, 0.77, and 1.15. These values are compatible with the remark in Section~\ref{sec:accuracy} that MASE values around 1 or lower indicate an excellent accuracy. The test MAPE values for the three curves are 27\%, 7\%, and 8\%. 
The right-hand plot 
shows the evolution of $\bar{S}(t)$.
}
\label{fig:SHR_12PA_BEL_traintstart8_traintstop38_c111}
\end{figure}

Figure~\ref{fig:SHR_12PA_BEL_traintstart8_traintstop38_c111}\footnote{\label{fn:results-BEL-peak}Reproduce with SHR\_19PA\_py\_BELsum\_1sttraintstart16\_1sttraintend32\_1sttesttend123\_c111.zip} compares the observed hospitalization curve with the hospitalizations forecasted by the model for a few train periods chosen around the first peak. The forecast errors are indeed remarkably small. This finding is compatible with the considerations in Section~\ref{sec:b-then-S0}, where we concluded that the estimation of $\bar\beta$, under a simple noise model, should be most accurate slightly before the peak of $H_o(t)$. 

\begin{figure}[t]
\centerline{\includegraphics[width=.9\textwidth]{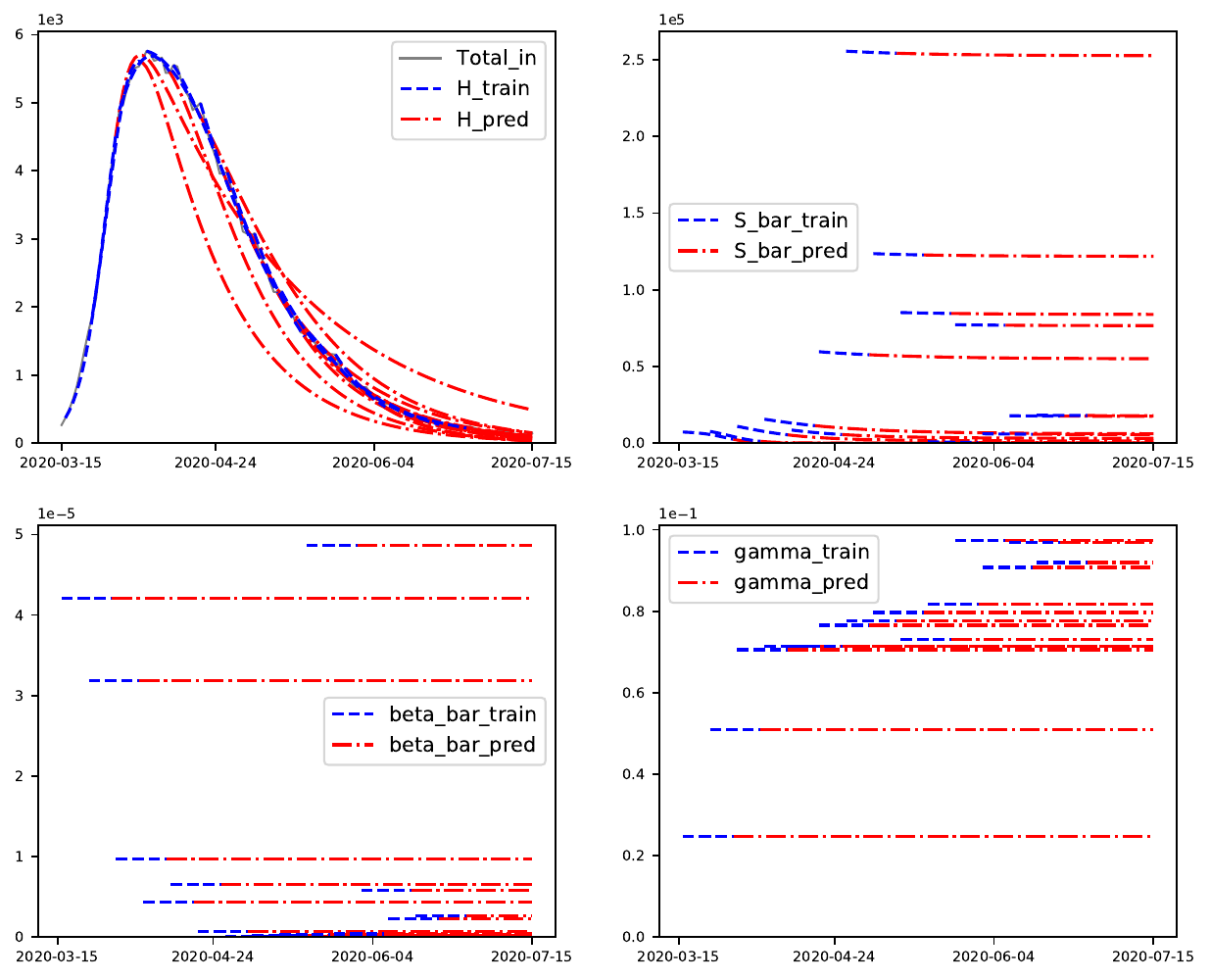}}
\caption{Belgium, various train periods.}
\label{fig:SHR_15PA_py_BEL_1sttraintstart1_1sttraintstop15_c111_2D}
\end{figure}

In Figure~\ref{fig:SHR_15PA_py_BEL_1sttraintstart1_1sttraintstop15_c111_2D},\footnote{Reproduce with SHR\_19PA\_py\_BELsum\_1sttraintstart1\_1sttraintend15\_1sttesttend123\_c111.zip} we superpose the results obtained with various train periods of 14 days for Belgium. The figure further corroborates the comments of Section~\ref{sec:b-then-S0}. In particular, if the train period is fully located before the peak, then the forecasts are rather inaccurate. Placing the train period around the peak gives excellent forecast results. When the train period is fully located in the decrease phase, the estimation of $\bar\beta$ and $\bar{S}(t_i)$ is seen to be very sensitive, but this does not affect much the quality of the forecast of $H(t)$.

\begin{figure}[t]
\centerline{\includegraphics[width=.9\textwidth]{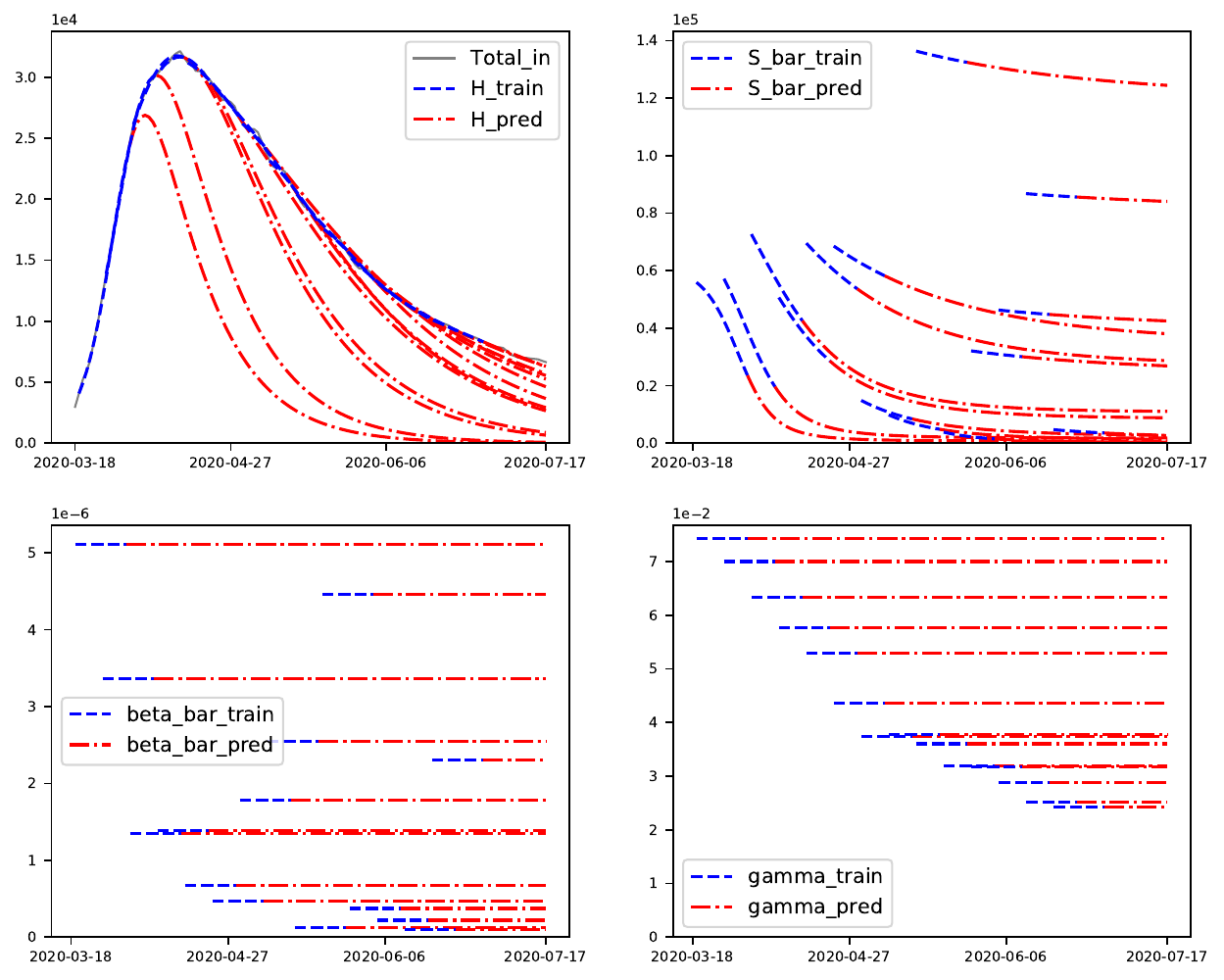}}
\caption{France, various train periods.}
\label{fig:SHR_15PA_py_FRA_1sttraintstart1_1sttraintstop15_c111_2D}
\end{figure}

Figure~\ref{fig:SHR_15PA_py_FRA_1sttraintstart1_1sttraintstop15_c111_2D}\footnote{Reproduce with SHR\_19PA\_py\_FRAsum\_1sttraintstart1\_1sttraintend15\_1sttesttend122\_c111.zip} is the equivalent of Figure~\ref{fig:SHR_15PA_py_BEL_1sttraintstart1_1sttraintstop15_c111_2D} for France. Again, the experiments are compatible with the comments of Section~\ref{sec:b-then-S0}. In experiments not reported here, we also considered some departments separately, with fairly similar results. 

A disconcerting aspect is the evolution of the estimated $\gamma$ as a function of the location of the train period. In Figure~\ref{fig:SHR_15PA_py_BEL_1sttraintstart1_1sttraintstop15_c111_2D} (Belgium), the estimation of $\gamma$ is grouped around 0.08 for several train periods. However, in Figure~\ref{fig:SHR_15PA_py_FRA_1sttraintstart1_1sttraintstop15_c111_2D} (France), the estimation of $\gamma$ keeps decreasing, indicating that the daily number of patients leaving the hospital is an increasingly small fraction of the number of patients at the hospital. In view of~\eqref{eq:gamma}, this casts doubt on the $L_o$ values (see Section~\ref{sec:data}).

Since, for some train periods, highly accurate two-months-ahead forecasts are obtained with the SH model~\eqref{eq:SH-DT}, whose order is as low as two, we also considered the forecasts obtained by an even simpler model, namely an exponential decay (or growth if $\gamma<0$) of the form $H(t) = H(t_i) e^{-\gamma(t-t_i)}$. The two parameters $H(t_i)$ and $\gamma$ are estimated in the least-square sense over the train period. Figure~\ref{fig:SHR_22PA_py_BELsum_1sttraintstart1_1sttraintend15_1sttesttend75_c111_MASE} shows that, for most train periods, the forecast errors obtained with the exponential model and the SH model are fairly similar. However, the SH model produces much more accurate forecasts than the exponential model when the train period is located around the peak.

\subsection{Forecasts from an automatically chosen region around the peak}
\label{sec:results-peak-auto}

In this section, we investigate if the accurate forecasts obtained when the train period is around the first peak are also observed in other datasets.

In order to conduct experiments on a large number of datasets, we have to automatize the selection of the train period around the peak. Moreover, we want the procedure
to be free of data leakage, i.e., it has to select the end time $t_c$ of the train period without ever reading the data for $t>t_c$. 

The proposed train-period selection procedure goes as follows. We first select a time span $N$; in our experiments, we choose $N=7$. In order to robustify the procedure, we smooth out the $H_o$ time series by computing its moving average (MA):
\begin{equation}  \label{eq:MA}
H_o^{\text{MA}}(t) = \text{mean}(H_o(t-N),\dots,H_o(t+N)).
\end{equation}
Then we compute $\hat{t} = \min\{t: \argmax_{\tau\leq t}H_o^{\text{MA}}(\tau) = t-N\}$; in other words, the highest value of $(H_o^{\text{MA}}(t))_{t\leq\hat{t}}$ is at $t=\hat{t}-N$, which we declare to be the location of the peak. This $\hat{t}$ is computed by letting $t$ increase until the argmax condition is satisfied. In view of~\eqref{eq:MA}, the data for $t>\hat{t}+N$ remains unseen, hence we choose $t_c := \hat{t}+N$ as the end time of the train period, thereby avoiding data leakage. The start of the train period is chosen as $t_i = \hat{t}-2N$.

The test accuracies are measured over the 60-day period $[t_c+1,\min(t_c+60,t_e)]$.

The results obtained for each district (province in Belgium and department in France) are summarized in Table~\ref{tab:results-peak-auto-BEL}\footnote{Reproduce with SHR\_22PA\_py\_BELeach\_1sttraintstart19\_1sttraintend41\_1sttesttend101\_c111.zip} (Belgium) and Table~\ref{tab:results-peak-auto-FRA}\footnote{Reproduce with SHR\_22PA\_py\_FRAeach\_1sttraintstart50\_1sttraintend72\_1sttesttend132\_c111.zip} (France). Moreover, for the whole of Belgium, we obtain the excellent MASE\_test = 0.70,\footnote{Reproduce with SHR\_22PA\_py\_BELsum\_1sttraintstart17\_1sttraintend39\_1sttesttend99\_c111.zip.}
and for the whole of France, we get a considerably poorer MASE\_test = 14.07.\footnote{Reproduce with SHR\_22PA\_py\_FRAsum\_1sttraintstart20\_1sttraintend42\_1sttesttend102\_c111.zip.}

The tables reveal that none of the districts admits an MASE\_test as low as the 0.70 obtained for the whole of Belgium. However, Section~\ref{sec:all-train-periods} has shown that an MASE\_test around 3 is still appreciably low. In view of Table~\ref{tab:results-peak-auto-BEL}, more than half of the Belgian provinces have an MASE\_test below 3.15. Table~\ref{tab:results-peak-auto-FRA} indicates that an MASE\_test below 3 occurs for at least 10\% of the French departments.

These experiments thus indicate that the SH model~\eqref{eq:SH-DT}, trained by the procedure described in Sections~\ref{sec:H0}--\ref{sec:b-S0}, can be an asset for planning health care resources and the easing of lockdown restrictions over the two--three forthcoming months as soon as the peak has been reached.

\begin{table}  
\caption{Statistics of forecast errors over all Belgian provinces for a train period automatically selected around the first peak. The left column specifies the various accuracy measures, as defined in Section~\ref{sec:accuracy}. The notation $P_n$ stands for the $n$th percentile.} \label{tab:results-peak-auto-BEL}
\begin{center}
\begin{tabular}{c|ccccccc}
\hline
Percentiles & min & $P_{10}$ & $P_{25}$ & $P_{50}$ & $P_{75}$ & $P_{90}$ & max \\ \hline
RRSE\_train & 0.30 & 0.32 & 0.33 & 0.33 & 0.45 & 0.55 & 0.65 \\
RRSE\_test & 0.13 & 0.16 & 0.20 & 0.32 & 0.36 & 0.40 & 0.99 \\
RRSE\_test/RRSE\_train & 0.24 & 0.53 & 0.63 & 0.79 & 0.92 & 1.15 & 1.52 \\
\hline
MASE & 0.61 & 0.79 & 1.18 & 1.43 & 2.47 & 2.60 & 8.89 \\
MASE\_train & 0.65 & 0.71 & 0.76 & 0.88 & 0.94 & 1.07 & 1.28 \\
MASE\_test & 1.51 & 1.81 & 2.23 & 3.15 & 3.87 & 4.73 & 14.90 \\
MASE\_test/MASE\_train & 1.54 & 2.21 & 2.57 & 3.69 & 4.78 & 6.08 & 16.62 \\
\hline
sMAPE\_train & 0.02 & 0.02 & 0.02 & 0.03 & 0.03 & 0.04 & 0.06 \\
sMAPE\_test & 0.09 & 0.11 & 0.20 & 0.22 & 0.47 & 0.49 & 0.83 \\
sMAPE\_test/sMAPE\_train & 4.31 & 5.20 & 6.44 & 8.00 & 14.60 & 20.01 & 35.69 \\
\hline
\end{tabular}
\end{center}
\end{table}

\begin{table}  
\caption{Same as Table~\ref{tab:results-peak-auto-BEL}, now for the French departments.} \label{tab:results-peak-auto-FRA}
\begin{center}
\begin{tabular}{c|ccccccc}
\hline
Percentiles & min & $P_{10}$ & $P_{25}$ & $P_{50}$ & $P_{75}$ & $P_{90}$ & max \\ \hline
RRSE\_train & 0.17 & 0.35 & 0.40 & 0.52 & 0.67 & 0.86 & 1.26 \\
RRSE\_test & 0.13 & 0.31 & 0.45 & 0.77 & 1.34 & 2.12 & 5.79 \\
RRSE\_test/RRSE\_train & 0.33 & 0.50 & 0.77 & 1.38 & 2.95 & 4.43 & 11.85 \\
\hline
MASE & 0.37 & 1.26 & 2.12 & 4.16 & 7.77 & 11.43 & 23.46 \\
MASE\_train & 0.50 & 0.75 & 0.87 & 0.99 & 1.20 & 1.39 & 1.89 \\
MASE\_test & 1.88 & 2.95 & 4.37 & 8.50 & 14.08 & 18.83 & 90.10 \\
MASE\_test/MASE\_train & 2.05 & 2.83 & 4.23 & 7.66 & 16.15 & 21.14 & 68.08 \\
\hline
sMAPE\_train & 0.01 & 0.01 & 0.02 & 0.04 & 0.05 & 0.08 & 0.36 \\
sMAPE\_test & 0.06 & 0.18 & 0.26 & 0.45 & 0.79 & 1.09 & 1.80 \\
sMAPE\_test/sMAPE\_train & 1.33 & 4.17 & 6.94 & 12.88 & 23.25 & 33.34 & 56.45 \\
\hline
\end{tabular}
\end{center}
\end{table}

\subsection{Challenges}
\label{sec:challenges}

The MASE curves in Figure~\ref{fig:SHR_22PA_py_BELsum_1sttraintstart1_1sttraintend15_1sttesttend75_c111_MASE} have a rather jagged shape. For the exponential model, this can be attributed to the weekly variations of the hospitalization curve due to the fact that fewer patients are discharged during the weekend. For the SH model, the phenomenon is more pronounced because, more than the exponential model, it is able to fit the technical weekly variations in the data, resulting in poorer forecast accuracies for some train periods. A preliminary filtering aiming at reducing these technical weekly variations might lead to improved forecasts.

\begin{figure}[t]
\centerline{
\includegraphics[width=.7\textwidth]{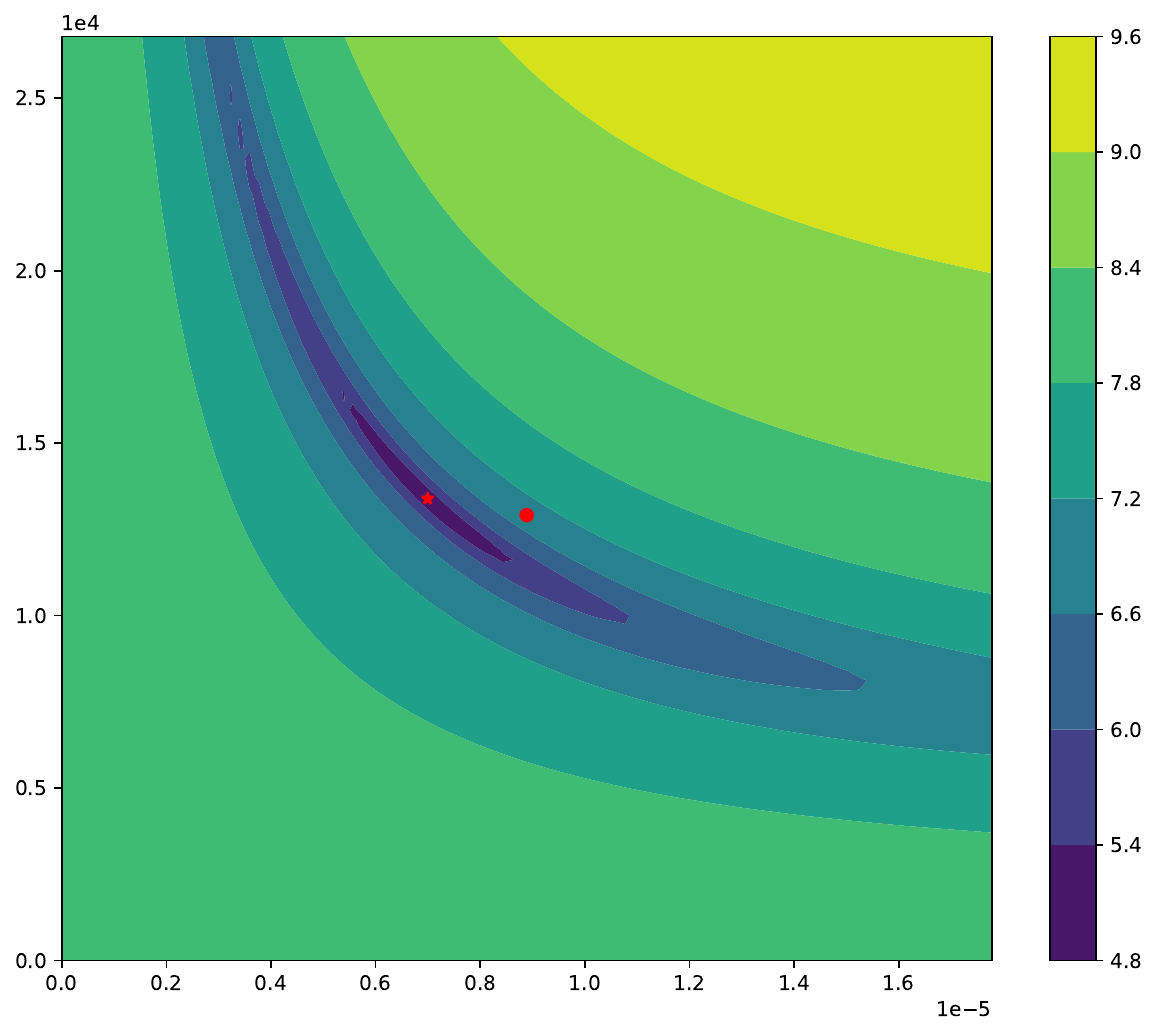}}
\caption{Belgium, contour plot of the objective function~\eqref{eq:phi} for a train period around the peak. Horizontal axis: $\bar\beta$; vertical axis: $\bar{S}(t_i)$. In order to make the minimizer easier to visualize, the plot shows equispaced-level curves of $\log(\phi - 0.90\,\phi_*$), where $\phi_*$ is the approximation of the minimal value of $\phi$ provided by the optimization solver. The red circle shows the initial guess obtained from Section~\ref{sec:b-then-S0} and the red star shows the approximate minimizer returned by the optimization solver. In our experiments, the optimization solver is scipy.optimize.fmin with its default parameters.}
\label{fig:SHR_19PA_py_BELsum_1sttraintstart16_1sttraintend32_1sttesttend123_c111_contour}
\end{figure}

Another cause, specific to the SH model, is that the objective function~\eqref{eq:phi}, which has to be minimized to find the estimands $\bar\beta$ and $\bar{S}(t_i)$, has a narrow valley, as illustrated in Figure~\ref{fig:SHR_19PA_py_BELsum_1sttraintstart16_1sttraintend32_1sttesttend123_c111_contour}.\footnote{Reproduce with SHR\_22PA\_py\_BELsum\_1sttraintstart17\_1sttraintend39\_1sttesttend99\_c111.zip} This makes it challenging to accurately compute the minimizer. The situation is worst for periods where the data is accurately fitted by the exponential model, because this means that an accurate fit is obtained with $\bar{S}(t)$ almost constant in~\eqref{eq:SH-DT-H}, which makes it difficult to separately estimate $\bar\beta$ and $\bar{S}(t_i)$. For some train periods, the optimization solver scipy.optimize.fmin terminates far away from the global minimum of the objective function, resulting in a significantly suboptimal fit. Developing optimization methods tailored to the specific landscape of the objective function~\eqref{eq:phi} is an important topic for further research.

\section{Conclusion}  
\label{sec:conclusion}

We have assessed the accuracy of COVID-19 hospitalization forecasts obtained with the SH model~\eqref{eq:SH-DT}, a simple discrete-time dynamical system with only two state variables and two (time-independent) parameters.
The experiments in Section~\ref{sec:results} have shown that the proposed method has a remarkably good fitting accuracy over the whole first wave. It also produces remarkably accurate forecasts over certain time ranges for some areas (Belgium, some Belgian provinces, and a few French departments). 

However, there are also time ranges and areas where the forecasts are very inaccurate. In particular, when it is trained before the peak, the model produces rather poor forecasts for the position and height of the peak and for the subsequent decrease. The model is also unable to produce multiple peaks in order to fit or forecast rebounds. The forecasts returned by the model should thus be taken with much caution. 

Another source of caution is that, even though it requires to estimate only four (time-invariant) estimands, fitting the SH model~\eqref{eq:SH-DT} to the data is already not a trivial task. We have not ruled out
the situation where the considered objective function would be multimodal. The optimization solver might thus get stuck in a local nonglobal minimum, yielding a suboptimal fit of the train data and possibly poorer forecasts than what the global minimum would achieve. Moreover, even if the objective function is unimodal, the stopping criterion of the solver may trigger before a suitably accurate approximation of the minimum is reached, as discussed in Section~\ref{sec:challenges}.

If the proposed model is used to guide prevention policies, then further caveats are in order. We have seen that the estimation of $\bar\beta$ is very sensitive. Hence the proposed model can hardly help assess the impact of prevention measures on $\bar\beta$. Without knowing sufficiently accurately the impact of prevention measures on $\bar\beta$, we may not aptly use the model to predict their impact on  the evolution of the hospitalizations.

Yet another caveat is that it may be tempting to deduce from the excellent fit with a constant-parameter model (Figure~\ref{fig:SHR_12PA_BEL_traintstart1_traintstop117_c100_4Dopt}) that the evolution of the prevention measures over the dataset period has had no impact on $\bar\beta$. But the deduction is flawed. Indeed, in view of the comments made in Section~\ref{sec:b-then-S0}, the available data could also be very well explained with fairly large jumps in $\bar\beta$ during the decrease phase. 

In spite of all these caveats, the hospitalization forecasts returned by the method, and also the evolution of $\bar{S}(t)$, might be of practical use in the context of various disease outbreaks, e.g., for resource planning. To this end, it will be important to understand which specific features of the COVID-19 outbreak in Belgium made it possible to forecast so accurately the hospitalization decrease several months ahead.

\section*{Acknowledgement}

This work benefited from discussions with several colleagues, in particular Fr\'ed\'eric Crevecoeur, Pierre Dupont, Alexey Medvedev, Pierre Schaus, and Lo\"{\i}c Van Hoorebeeck.

\bibliographystyle{alphaurl}

\bibliography{bibfile}


\end{document}